\documentclass[prl,superscriptaddress,twocolumn,nopacs,letter,10pt]{revtex4-2}
\usepackage{graphicx,bm,times}
\usepackage{amsmath}
\usepackage{amsfonts}
\usepackage{amssymb}
\usepackage{color}
\usepackage{hyperref}
\hypersetup{
	colorlinks = true,
	allcolors = {blue}
}

\usepackage[utf8]{inputenc}
\usepackage[T1]{fontenc}
\usepackage{amsmath,amssymb}
\usepackage{lineno}
\usepackage{float}
\usepackage{siunitx}
\usepackage{soul}
\usepackage{xfrac}

\newcommand{\threebythree}{$3\!\times{}\!3$ }
\newcommand{\sqrtthirteen}{$\sqrt{13}\!\times{}\!\sqrt{13} R(13.9^\circ)$}

\begin{document}
\title{Charge transfer empties the flat band in 4H\textsubscript{b}-TaS\textsubscript{2} - except at the surface}

\author{Mihir Date}

\email{mihir.date@mpi-halle.mpg.de, matthew.watson@diamond.ac.uk}

\affiliation {{Diamond Light Source Ltd, Harwell Science and Innovation Campus, Didcot, OX11 0DE, UK}}
\affiliation {Max Planck Institute of Microstructure Physics, Weinberg 2, Halle (Saale), Germany}

\author{Hyeonhu Bae}
\affiliation{{Weizmann Institute of Science, 234 Herzl Street, POB 26, Rehovot 7610001 Israel}}

\author{Alex Louat}
\affiliation {{Diamond Light Source Ltd, Harwell Science and Innovation Campus, Didcot, OX11 0DE, UK}}

\author{Gabriele Domaine}
\affiliation {Max Planck Institute of Microstructure Physics, Weinberg 2, Halle (Saale), Germany}

\author{Niels B. M. Schr\"{o}ter}
\affiliation {Max Planck Institute of Microstructure Physics, Weinberg 2, Halle (Saale), Germany}

\author{Enrico Da Como}
\affiliation {Department of Physics, Condensed Matter and Quantum Materials Group, University of Bath, Claverton Down, Bath BA2 7AY, United Kingdom}

\author{Binghai Yan}
\affiliation{{Weizmann Institute of Science, 234 Herzl Street, POB 26, Rehovot 7610001 Israel}}

\author{Matthew D. Watson}
\affiliation {{Diamond Light Source Ltd, Harwell Science and Innovation Campus, Didcot, OX11 0DE, UK}}

\begin{abstract}
The 4H\textsubscript{b} polytype of TaS\textsubscript{2} is a natural heterostructure of H and T-type layers. Intriguing recent evidence points towards a possibly chiral superconducting ground state, unlike the superconductivity found in other polytypes where the T layers are absent, requiring understanding of the possible contributions of electrons from the T layers. Here we use micro-focused angle resolved photoemission spectroscopy to reveal that the T termination of the 4H\textsubscript{b} structure is metallic, but a subsurface T layer - seen below an H termination and thus more representative of the bulk case - is gapped. The results imply a complete charge transfer of 1 electron per 13 Ta from the T to adjacent H layers in the bulk, but an incomplete charge transfer at the T termination, yielding a metallic Fermi surface with a planar-chiral character. A similar metallic state is found in an anomalous region with likely T-H-H' stacking at the surface. Our results exclude cluster Mott localisation in either the bulk or surface of 4H\textsubscript{b}-TaS\textsubscript{2} and point to a scenario of superconductivity arising from Josephson-like tunneling between the H layers. 
\end{abstract}

\date{\today}

\maketitle

\section{Introduction}

The Ta-dichalcogenides offer a uniquely tunable platform for exploring the archetypal phenomena of correlated electron systems: charge density waves (CDWs), Mott insulating behaviour, and superconductivity, driven by electron-electron and electron-phonon interactions. The 4H\textsubscript{b} polytype of TaS\textsubscript{2} weaves these strands together in a natural heterostructure of T and H layers: the T layers undergo a commensurate charge density wave, while the H layers exhibit a metallic Fermi surface that presumably yields the principal density of states requisite for the emergence of superconductivity at low temperatures. With a $T_c$ of 2.7 K, the superconducting state is intriguing: there is evidence for gapless excitations at low temperatures~\cite{ribak_chiral_2020, persky_magnetic_2022, dentelski_robust_2021, wang_evidence_2024}, and a spontaneous jump in the zero field muon spin relaxation rate at $T_c$ suggests that the ground state could host a chiral superconducting order parameter \cite{ribak_chiral_2020}. A two-component order, parametrized by chiral and nematic phases, has also been proposed \cite{nayak_evidence_2021}. Further evidence that time-reversal symmetry may be broken in the superconducting phase comes from the existence of vortices in zero applied magnetic field \cite{persky_magnetic_2022,liu_magnetization_2024} and a $\pi$ shift in the Little-Parks effect \cite{almoalem_observation_2024,fischer_mechanism_2023}. 

While several dichalcogenides in the 2H structure are superconductors, the apparently unique nature of the superconductivity in the 4H\textsubscript{b}-TaS\textsubscript{2}  has drawn theoretical attention to the role of the T layers. The divergence in theoretical treatments of these, however, underlines both their complexity and the lack of experimental evidence. The \sqrtthirteen{} commensurate charge density wave (abbreviated as $\sqrt{13}$ and CCDW hereafter), would, in the case of a freestanding T layer, localize a single electron of over a cluster of 13 Ta, which arrange in a Star-of-David formation  below 315 K \cite{wilson_charge-density_1975}. In the absence of electron-electron interactions, this remaining electron would lead to a half-filled metallic band with narrow bandwidth - the "flat band" - due to geometrically unfavourable in-plane hopping \cite{rossnagel_origin_2011}. However in the bulk 1T-TaS\textsubscript{2}, no such flat band exists at the Fermi level ($E_F$), and instead the observation of apparent energy gaps in tunneling and spectroscopy \cite{kim_observation_1994,perfetti_unexpected_2005,wang_band_2020,fei_understanding_2022}, has led to suggestions of a cluster Mott insulator and emergent quantum spin liquid state \cite{law_1t-tas_2_2017,he_spinon_2018}. 

Carrying the idea of cluster Mott localisation across to the 4H\textsubscript{b} case seems compelling, especially as the complication of interlayer hopping is suppressed \cite{ritschel_stacking-driven_2018}. It was initially suggested that the chiral superconductivity may be promoted by spin fluctuations arising from Mott-like correlations in T layer \cite{ribak_chiral_2020}, and some subsequent theoretical approaches similarly rely on magnetic moments or fluctuations in the T layers \cite{dentelski_robust_2021,luo_is_2024,lin_kondo_2024}. However, an important difference in the 4H\textsubscript{b} polytype is that the H layers have a higher work function, enabling transfer of electrons from the T to the H layers \cite{sanchez-ramirez_charge_2024,almoalem_charge_2024}. This charge transfer has been also explored in the equally intriguing context of H-T bilayers \cite{vano_artificial_2021, ayani_electron_2024}, where some calculations suggest a charge transfer of 1 $e^-$ per cluster of 13 Ta, completely depopulating the flat band \cite{crippa_heavy_2024}. If this were the case in bulk 4H\textsubscript{b}-TaS\textsubscript{2}, a different perspective on the superconductivity would be required, based instead on a Josephson-like scenario \cite{fischer_mechanism_2023}. 

%% Fig 1
\begin{figure*}[ht]
\centering
\includegraphics[width=280pt, angle=-90]{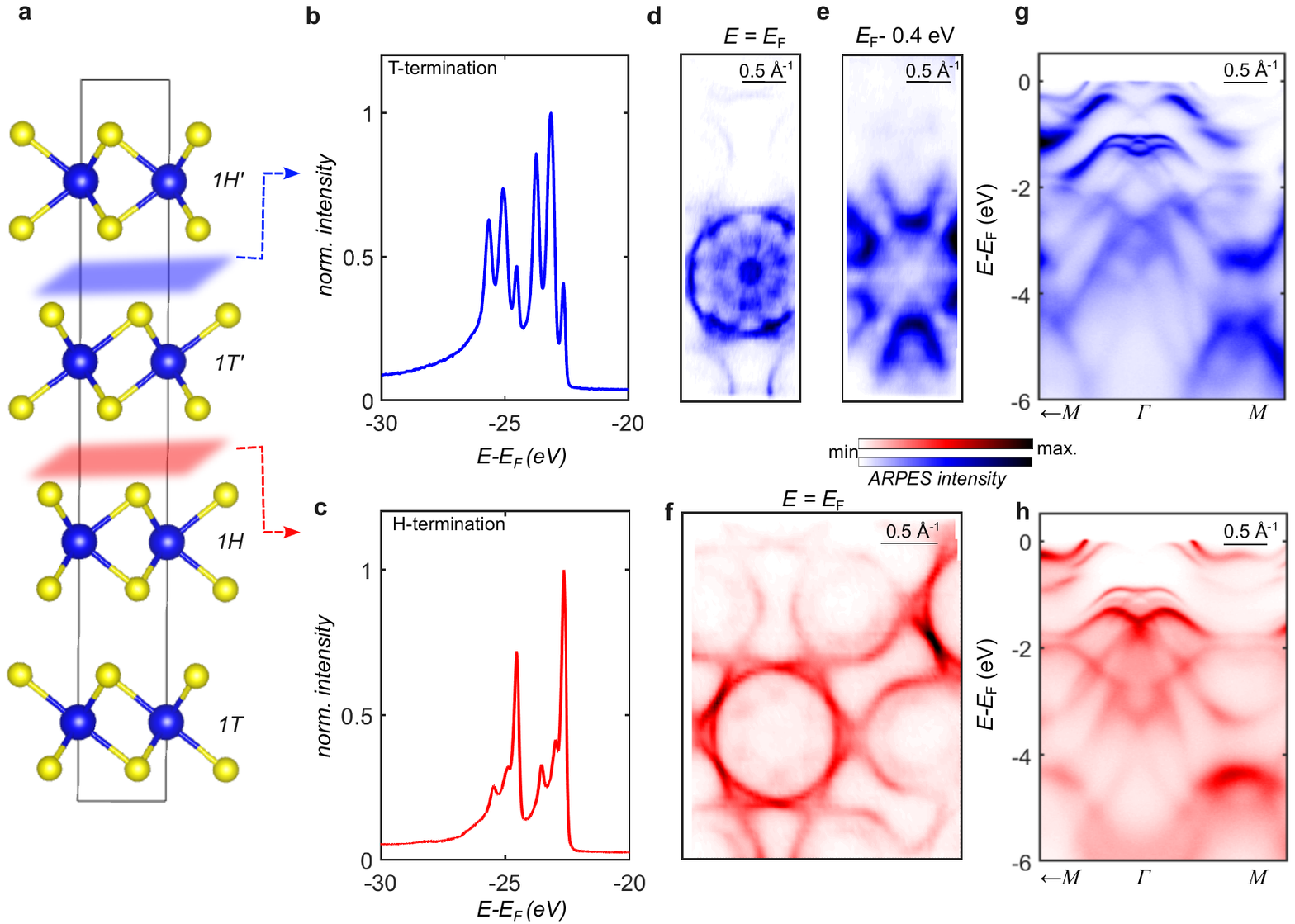}
\caption{\textbf{Termination-dependent ARPES on 4H\textsubscript{b}-TaS\textsubscript{2}.} \textbf{(a)} The crystal structure of 4H\textsubscript{b}-TaS\textsubscript{2}, as viewed down the $a$ axis; black rectangle is the unit cell. Two possible cleavage planes, indicated in blue and red, produce T and H terminated surfaces, respectively. Ta $4f$ shallow core levels on the \textbf{(b)} T-and \textbf{(c)} H-terminated surfaces. Isoenergy surface of the T-terminated 4H\textsubscript{b}-TaS\textsubscript{2} at \textbf{(d)} the Fermi level, and \textbf{(e)} E$_F$-0.4 eV. \textbf{(d)} Dispersion along the $\mathrm{M\Gamma{}M}$ direction on the T-termination. The equivalent Fermi surface and dispersion on the H-terminated surface are shown in \textbf{(f)} and \textbf{(h)}.}
\label{Fig1}
\end{figure*}

Despite the wealth of scanning tunneling microscopy (STM) measurements, the lack of momentum-resolved information obscures the full picture of the band filling, leading to a variety of scenarios being proposed for both H-T bilayers \cite{vano_artificial_2021, wan_evidence_2023, ayani_electron_2024,crippa_heavy_2024} and bulk 4H\textsubscript{b}-TaS\textsubscript{2} \cite{ekvall_atomic_1997}. Angle-resolved photoemission spectroscopy (ARPES) is the perfect tool to examine the filling of the T-derived flat bands \cite{ribak_chiral_2020,almoalem_charge_2024,pudelko_probing_2024,yang_signature_2025}. There are, however, two important subtleties in measurements on 4H\textsubscript{b}-TaS\textsubscript{2}: the termination-dependence, since in principle both H and T-like terminations are possible, and the distinction between bulk band structure and surface-confined states. 

Here we propose a mechanism of charge transfer in 4H\textsubscript{b}-TaS\textsubscript{2} based on our high-resolution and termination-specific micro-ARPES data. First, we introduce the electronic structures of T and H terminated 4H\textsubscript{b}-TaS\textsubscript{2}, in particular highlighting the ``planar chirality" in the former. Together with layer-resolved \textit{ab initio} calculations, we show that the dispersion of bulk and surface T bands differ substantially. Importantly, we find that the subsurface T layer, below the H termination, is gapped, with dispersions significantly shifted compared to those on the T termination. This is direct evidence that, in the bulk of 4H\textsubscript{b}-TaS\textsubscript{2}, there is complete charge transfer of 1 $e^-$ per 13 Ta. Finally, we discuss the electronic structure of a region with anomalous stacking, plausibly 1T/1H/1H', where we similarly find a metallic T surface with slightly increased filling. Our results leave no room for electronic correlations either at the surface or in the bulk of 4H\textsubscript{b}-TaS\textsubscript{2}, excluding theoretical explanations of its superconductivity that depend on either local moments or correlated electrons in the T layers.

\section{Results}

%% Fig 2 
\begin{figure*}[ht]
\centering
\includegraphics[width=280pt,angle=-90]{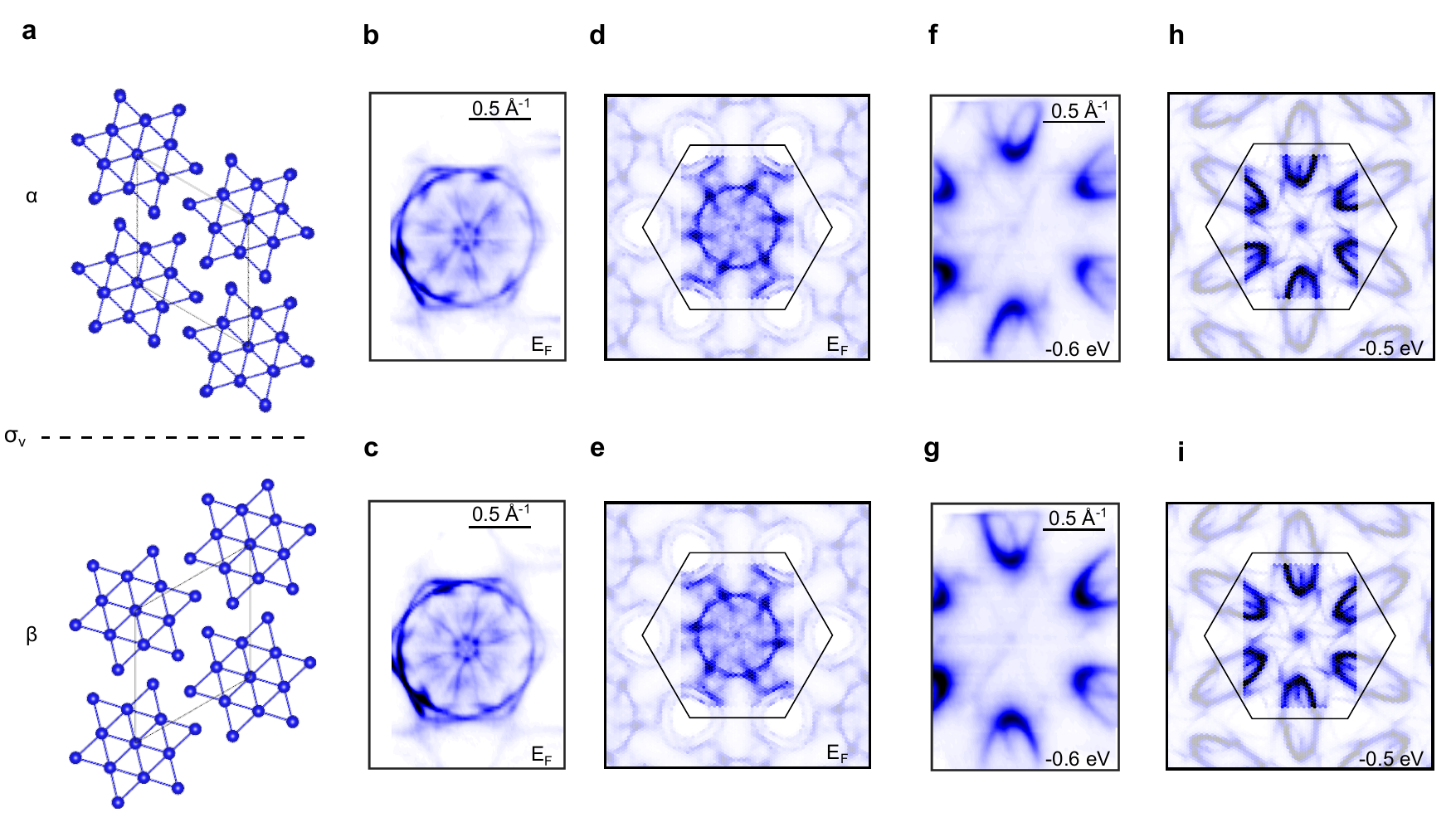}
\caption{\textbf{Planar chiral Fermi surface of 4H\textsubscript{b}-TaS\textsubscript{2}.} \textbf{ (a)} The real-space orientation of the SODs in the $\alpha$ (top) and $\beta$ (bottom) phases (Ta atoms only), and their respective experimental Fermi surfaces in \textbf{(b)} and \textbf{(c)}. The asymmetric ARPES intensity distribution is highlighted in the isoenergy slices at $E=E_F$-0.6~eV of the \textbf{(f)} $\alpha$ and \textbf{(g)} $\beta$ phases, respectively. \textbf{(d,e)} and \textbf{(h,i)} show DFT calculations at E$_F$ and $E=E_F$-0.6~eV for bilayer 1T/1H-TaS\textsubscript{2}, with the $\alpha$ and $\beta$ type $\sqrt{13}$ CDW in the T-layer.}
\label{Fig2}
\end{figure*}

%% Fig 1 text
The crystal structure of 4H\textsubscript{b}-TaS\textsubscript{2}, shown in Fig.~\ref{Fig1}(a), consists of a four-layered unit cell, which at temperatures above the onset of $\sqrt{13}$ CCDW at 315~K \cite{di_salvo_preparation_1973} is described in the centrosymmetric $P6_3/mmc$ space group (no.194) \cite{katzke_phase_2004}. The as-cleaved samples show both T and H-terminated surfaces, on length scales of approximately 20 $\mu{}$m, which conveniently allows us to measure the electronic structure of the singly terminated regions with the experimental beam spot of 4-5$\mu{}$m. 

The surface terminations are clearly distinguished by their characteristic Ta $4f$ core level spectra, as shown in Fig.~\ref{Fig1}(b,c). The T-terminated 4H\textsubscript{b}-TaS\textsubscript{2} shows six sharp peaks, or three within each of the $4f_{7/2}$ and $4f_{5/2}$ submanifolds. The most intense peaks are attributed to the inequivalent Ta sites in the $\sqrt{13}$ lattice reconstruction: within the 13-Ta clusters, the outer, middle and innermost Ta atoms show differing chemical shifts with  intensity ratio 6:6:1 \cite{hughes_site_1995,wolverson_first-principles_2022}, with the latter not contributing a clearly resolvable peak by itself. The two remaining peaks seen on the T termination, with the lowest binding energy, are ascribed to photoemission from the subsurface H layer. Contrastingly, on the H-terminated 4H\textsubscript{b}-TaS\textsubscript{2}, the intensity is strongest on the peaks at lowest binding energy, resembling the doublet structure of 2H-TaS\textsubscript{2} \cite{pudelko_probing_2024}, but again there is some additional structure from the subsurface T layers. Such contributions to the photoemission intensity from the subsurface becomes critical in our later results. 

On the H termination, the Fermi surface displayed in Fig.~\ref{Fig1}(f) shows a central hole pocket surrounded by a six-fold symmetric dogbone hole pocket. In unison with previous ARPES studies \cite{ribak_chiral_2020,almoalem_charge_2024}, we find no evidence of a \threebythree CDW order in the topmost 1H layer. Most likely this is because our measuring temperature of 30~K exceeds the transition temperature of 20~K \cite{di_salvo_preparation_1973}, associated with the weak \threebythree CDW seen by STM \cite{ekvall_atomic_1997,nayak_evidence_2021,bang_charge-ordered_2024,geng_correlated_2024}. The suppression of the  \threebythree CDW order distinguishes 4H\textsubscript{b}-TaS\textsubscript{2} from the 2H polytypes of both TaS\textsubscript{2} and TaSe\textsubscript{2}, as well as 4H\textsubscript{b}-TaSe\textsubscript{2}, in all of which clear signatures of \threebythree CDW order are observed by ARPES \cite{borisenko_pseudogap_2008,pudelko_probing_2024,watson_folded_2025}. In this sense the Fermi surface measured on the H termination is quite simple, and varies from the (unreconstructed) 2H-TaS\textsubscript{2} by a small electron doping \cite{almoalem_charge_2024} and the apparent band degeneracy (i.e. a very small and unresolved splitting) seen along $\mathrm{\Gamma{}M}$ due to the suppression of interlayer hopping.   

In comparison, the electronic structure of the T-terminated 4H\textsubscript{b}-TaS\textsubscript{2} is rather complex, and will be the focus of this paper. The first key message is that the T-terminated surface is metallic,  with a Fermi surface as shown in Fig.~\ref{Fig1}(d). There is a spectrally intense $\Gamma$-centered pocket, from which thin petal-like bands emanate. These features are distinct from the weaker intensity from the from the subsurface metallic H-bands. At a higher binding energy of -0.4~eV, we find sections of partially gapped ``cigar” shaped pockets, resembling sections of the expected cigar-shaped Fermi surfaces of the parent 1T phase \cite{rossnagel_origin_2011}. In the corresponding ARPES dispersion measured along $\Gamma-\rm{M}$ shown in Fig.~\ref{Fig1}(g), the brightest features derive from the photoemission signal from the Ta $5d$ orbitals of the terminating T layer. The opening of energy gaps and band folding from the $\sqrt{13}$ CCDW is prominent, exhibiting many similarities to the 1T-TaS\textsubscript{2}. What is unique to the 4H\textsubscript{b}-TaS\textsubscript{2}, however, is that at $\Gamma$ there is a shallow occupation of a band at $E_F$ on the T termination in Fig.~\ref{Fig1}(g), though not on the H termination in in Fig.~\ref{Fig1}(h). This indicates a partial occupation of the flat band on the T termination of 4H\textsubscript{b}-TaS\textsubscript{2}.
  
%% ACTION MIHIR: Fig 3e, remove ticks and label as "Intensity"
%% Fig 3 
\begin{figure*}[ht]
\centering
\includegraphics[width=\textwidth]{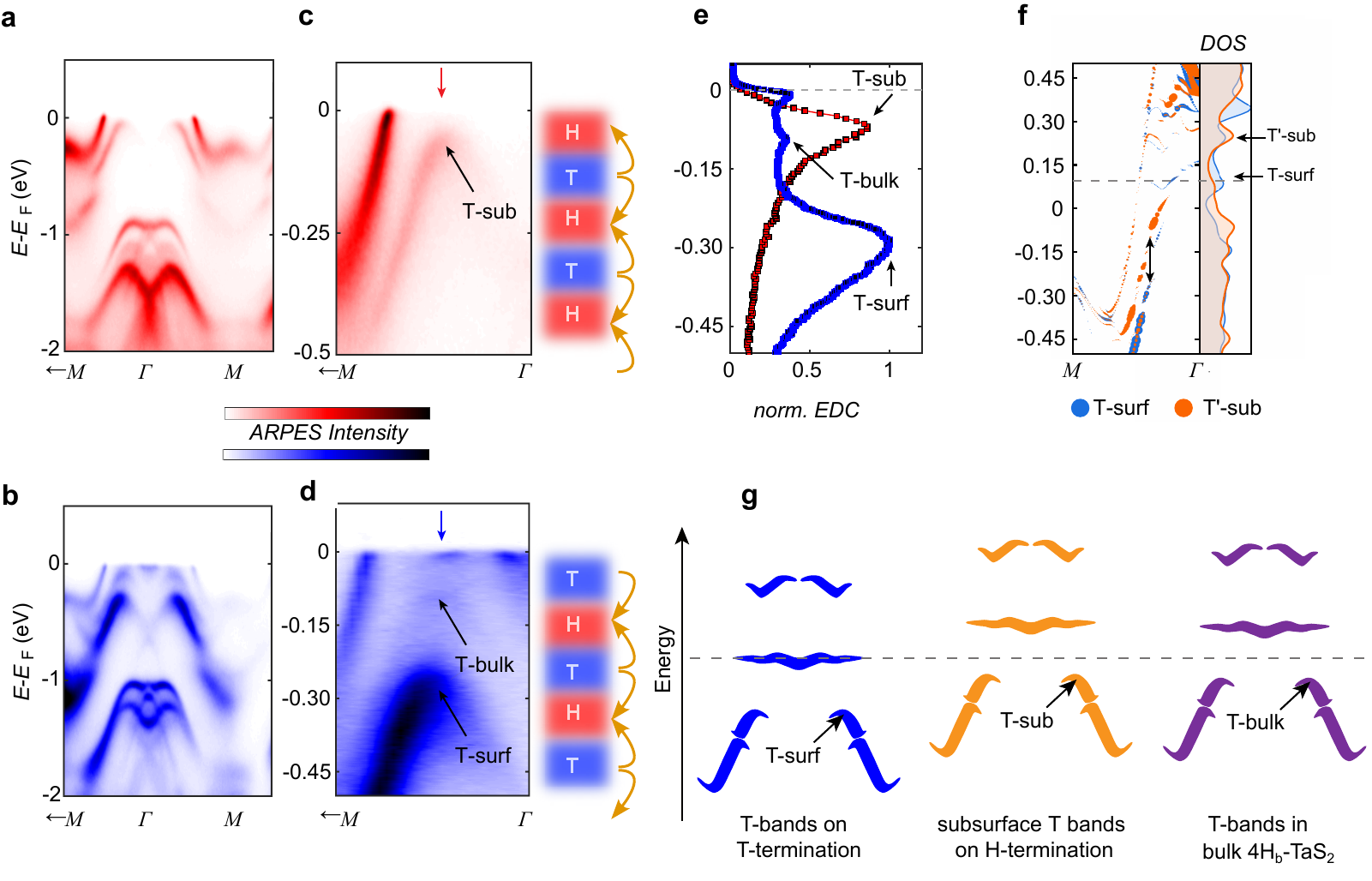}
\caption{\textbf{Differing charge transfer and band filling for the bulk and the surface.} Constant momentum cuts along the $\Gamma$-M direction on the \textbf{(a)} H and \textbf{(b)} T terminations measured with 100 eV photon energy. \textbf{(c)} and \textbf{(d)} Equivalent measurements at 50 eV, showing the low energy sector. The outset of \textbf{(c)} and \textbf{(d)} illustrates how unequal charge transfer at the surface and the bulk influences the band dispersion of H and T terminated 4H\textsubscript{b}-TaS\textsubscript{2}, respectively. \textbf{(e)} EDCs of the T-bands integrated over a $\pm 0.05$ {\AA}$^{-1}$ momentum window, as indicated by the red and blue arrows in \textbf{(c)} and \textbf{(d)}, respectively. \textbf{(f)} Layer-projected DFT bandstructure of a four-layer T/H/T'/H' slab. The energy separation between the surface (blue) and subsurface/bulk (orange) T bands is highlighted by the double-headed arrow. The indicated peaks in the density of states (DOS) corresponding to the flat bands also shows a similar separation. Note that the experimental filling on the T surface intersects the T-surf flat band, a little above the calculated $E_F$ of the slab, as indicated as indicated by the dashed gray line. \textbf{(g)} Schematic of spectral weight of the T-bands from the surface of the T-termination (T-surf), the subsurface below an H-termination (T-sub), and the bulk of 4H\textsubscript{b}-TaS\textsubscript{2}. }
\label{Fig3}
\end{figure*}

%% Fig 2 text
The intricate details of the metallic Fermi surface found on the T termination emerge in Fig.~\ref{Fig2}, measured with a lower photon energy and higher energy resolution (Methods) which reveal its intrinsic planar chirality. Phenomenologically, in the $\sqrt{13}$ reconstruction the in-plane mirror symmetries are broken, and therefore constant energy maps in the $k_x\!-\!k_y$ plane can exhibit planar chirality. Similar spectral features lacking mirror symmetry have been observed in 1T-TaS\textsubscript{2} (albeit not at $E_F$) \cite{ngankeu_quasi-one-dimensional_2017,ritschel_stacking-driven_2018,yang_visualization_2022,qi_temperature_2024,yang_signature_2025} and 4H\textsubscript{b}-TaSe\textsubscript{2} \cite{louat_pseudochiral_2024,watson_folded_2025}, and described using varying nomenclature, which we expand upon in the Supplementary Information (SI). Here, however, we can simply use the longstanding notation of $\alpha$ and $\beta$ domains, shown schematically in Fig.~\ref{Fig2}(a), where the in-plane lattice parameters of the $\sqrt{13}$ cell are rotated by $\pm{}13.9^\circ$; these domains correspond to the two different senses of rotation that are possible to observe in the Fermi maps. Unlike the 1T-TaS\textsubscript{2}, the $\alpha/\beta$ domains are small in the 4H\textsubscript{b}-TaS\textsubscript{2}, such that ARPES data most often appears to be mirror symmetric (e.g. Fig.~\ref{Fig1}(d) and in literature reports \cite{ribak_chiral_2020,pudelko_probing_2024}), which is now understood as a superposition of signal from both domains.

To support our interpretation, we compare our measurements with DFT calculations for a bilayer of 1T/1H-TaS\textsubscript{2}. The bilayer 1T/1H-TaS\textsubscript{2} structure is chosen to keep the calculations tractable and is the minimal to capture the essential CCDW physics and symmetries. We can clearly observe planar chiral spectral features of the projected T-layer states at the Fermi level in Figs.~2(d,e), as well as -0.6 eV (Figs.\ref{Fig2}(h,i)), both in good agreement with the experimental isoenergy slices. 

%%%%%%%%%

The observed Fermi surface on the T termination implies a partial occupation of the flat band. This would correspond, for example, to the calculations in Ref.~\cite{crippa_heavy_2024} for higher interlayer spacings, with a charge transfer greater than 0.4 but less than 1 $e^-$ per 13 Ta. Further evidencing the importance of charge transfer, we find the work functions of the H and T terminations of 4H\textsubscript{b}-TaS\textsubscript{2} to be 6.2 and 5.7 eV respectively (SI), close to the calculated values \cite{crippa_heavy_2024,sanchez-ramirez_charge_2024}. We note, however, that the T termination is adjacent to only one H layer, while in the bulk, or on the T subsurface under an H termination, the T layers are adjacent to two H layers. It is therefore not the case that the partial occupation of the flat band on the T termination is also representative of the bulk. 

To find fingerprints of a different charge transfer and band filling in the T subsurface, we reintroduce the energy dispersions measured on the two terminations in Fig.~\ref{Fig3}(a) and (b), focusing on the low energy sector. On the H termination in Fig.~\ref{Fig3}(a) we observe, in addition to the bright H-derived metallic bands, an inner and weaker band that not only resembles the bright state that dominates the T termination (Fig.~\ref{Fig3}(b)), but also bears the hallmarks of the $\sqrt{13}$ CCDW in the form of spectral gaps and backfolding. We can thus confidently ascribe this as the spectral weight of the subsurface T bands, but the entire bandwidth is shifted by 228 meV compared to the T termination. Exemplifying this, the band maximum of the bright T termination band marked as "T-surf" in Fig.~\ref{Fig3}(c) is at an energy of -297 meV whereas the equivalent from the subsurface T band, marked "T-sub." in Fig.~\ref{Fig3}(d) is at -69 meV (see EDCs in Fig.~\ref{Fig3}(e)).

\begin{figure*}[ht]
\centering
\includegraphics[width=300pt , angle=-90]{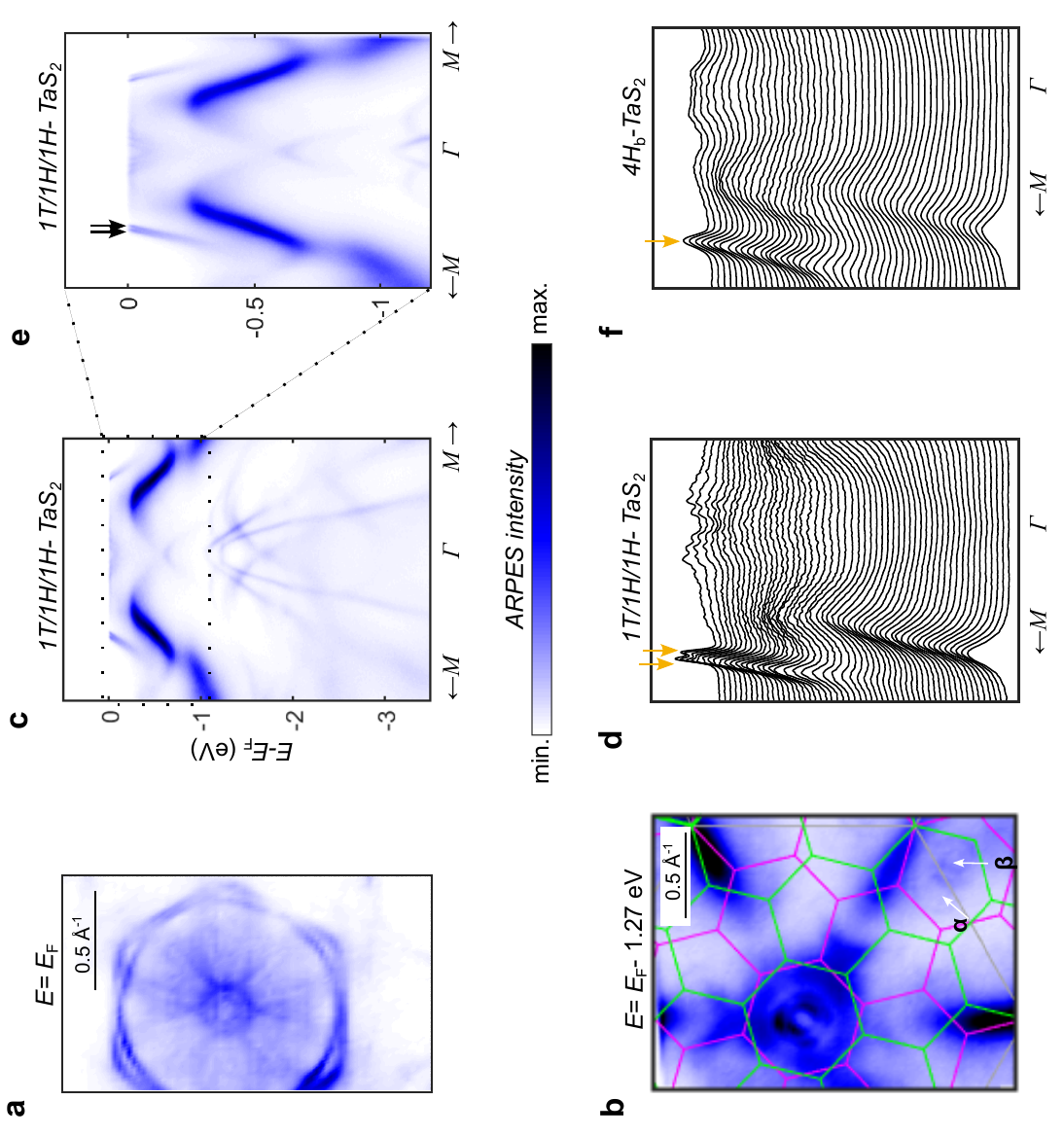}
\caption{\textbf{Electronic structure of an anomalously stacked 1T/1H/1H' region.} Isoenergy surfaces at \textbf{(a)} E$_F$ and \textbf{(b)} E$_F$- 1.27 eV. The green and magenta coloured hexagons show the Brillouin zones of the $\alpha$ and $\beta$ phases, respectively.\textbf{(c)} Dispersion along M-$\Gamma$-M on the T-termination with its closeup in \textbf{(e)} highlighting the T-bands around the $E_F$. The black arrows indicate two slightly split H-bands crossing the Fermi level. MDCs of the \textbf{(d)} anomalously stacked region and \textbf{(f)} normal 4H\textsubscript{b}-TaS\textsubscript{2}. The orange arrows in \textbf{(d)} show two MDC peaks representative of the two H-bands crossing E$_F$, whereas MDC of 4H\textsubscript{b}-TaS\textsubscript{2} shows one peak. Since this splitting resembles the bilayer splitting in 2H-TaS\textsubscript{2}, the measurement region most likely has two H-layers adjacent to the surface T-layer, resulting in a local T/H/H' stacking.}
\label{Fig4}
\end{figure*}

To corroborate our interpretation, in Fig.\ref{Fig3}(f) we consider a four-layered slab of alternating T and H layers, separated by 5.8 {\AA}; i.e. a single unit cell of the 4H\textsubscript{b} structure as depicted in Fig.~\ref{Fig1}(a). The layer-projected bandstructure displayed in Fig.~\ref{Fig3}(f) clearly highlights the energy separation between the blue (surface) and orange (subsurface) T-derived bands; the T-surf bands are lower by $\sim$ 200 meV. While there is a small discrepancy between the calculations and the experiments in the placement of the global $E_F$ of the system, the 4-layer calculations fully support, in both sign and magnitude, the intuitive concept that the charge transfer is less, and the band filling higher, on the surface than in the bulk. 

Crucially, there is no spectral weight at all at $E_F$ from the subsurface T in Fig.~\ref{Fig3}(a), implying that the flat band of the subsurface T layer is fully in the unoccupied states. The experimental scenario for the subsurface T band is therefore that it is insulating - but due to complete charge transfer, and not Mott physics. The question arises whether the subsurface is truly representative of the bulk. To answer this, we note that there is a very weak spectral feature measured on the T termination, labelled "T-bulk", where the equivalent band maximum is at -80 meV. This intensity plausibly comes from the third layer, and at this depth we can reasonably argue that any surface potential effects should be well screened and the band positions are closely representative of the bulk. Thus, both the subsurface and the bulk are in the same regime of complete charge transfer, but there is a small energy shift of 30 meV between them (see Fig.~\ref{Fig3}(e)). We summarise the experimental scenarios for the depth-dependent T-layer electronic structures with the schematic diagrams in Fig.~\ref{Fig3}(g).

%% Fig 4

While we have so far discussed T or H terminations of 4H\textsubscript{b}-TaS\textsubscript{2}, in fact these are not the only possibilities found experimentally. Hexagonal disulphides in general \cite{watson_novel_2024}, and crystals in the 4H\textsubscript{b} polytype in particular, are highly susceptible to stacking faults \cite{wilson_charge-density_1975}, thus there may be minority regions on the sample where the local stacking obeys alternative stacking orders \cite{zhang_kelvin_2023,geng_correlated_2024}. Fig.~\ref{Fig4}(a) shows the Fermi surface of a T-terminated region on the sample, which is noticeably different from the ones shown in Fig.~\ref{Fig2}. The difference is not simply in the restoration of mirror symmetry, which is easily understood by a superposition of $\alpha$ and $\beta$ domains, yielding backfolded S bands at $\Gamma$ points corresponding to both domains in Fig.~\ref{Fig4}(b). The filling of the T bands is higher than normal 4H\textsubscript{b}, as can be seen by comparing the Momentum Distribution Curves (MDCs) in Fig.~\ref{Fig4}(d,f). The feature labelled $T_{bulk}$ in Fig.~\ref{Fig3}(d) is absent here in Fig.~\ref{Fig4}(e), indicating that there may be not be T layers in the stack underneath the T termination. Additionally, the H bands are slightly split: such a band splitting along the $\Gamma$-M is a hallmark of interlayer hybridization, found also in trilayer calculations presented in the SI. Taken together, this suggests that the local stacking in this measurement region begins with 1T/1H/1H'.  

Three conclusions follow from this observation. First, we find that the T termination is also metallic here, with a band filling higher than that of normal 4H\textsubscript{b}, yet the band dispersions near $\mathrm{\Gamma}$ are quite steep and do not show any clear signatures of electron-electron interactions. Second, it is a reminder that physical crystals of 4H\textsubscript{b}-TaS\textsubscript{2} are typically heterogeneous and far from perfect \cite{wilson_charge-density_1975,bang_charge-ordered_2024}, and it may be important to consider the possible role of such anomalous stackings within the crystal when interpreting experimental data. Third, it exemplifies the value of micro-ARPES in identifying and measuring anomalous regions, which may be a route to stabilize the theoretically appealing T/T/H trilayer stacking \cite{bae_designing_2025}.

\section{Discussion}

With our prescription of a charge transfer mechanism in 4H\textsubscript{b}-TaS\textsubscript{2}, we have addressed the longstanding question of possible cluster Mott physics in T layers. We have shown that 4H\textsubscript{b}-TaS\textsubscript{2} hosts a metallic Fermi surface on the T termination, exhibiting planar chirality. However, this metallic state contains only tiny dot-like and strand-like spectral features, most likely corresponding to a filling much less than than 1 $e^-$ per 13 Ta, although a quantitative estimate was not possible due to its complexity. Since the putative Mott insulator would be principally applicable to the half-filled (1 $e^-$ per 13 Ta) case, our results do not strictly exclude Mott physics in a hypothetical freestanding and stoichiometric monolayer 1T-TaS\textsubscript{2}. In practice, however, monolayers are always grown or placed on a substrate of some kind, and our results show that differences in the work function can lead to charge transfer effects which can be comparable to, or even reach 1 $e^-$ per 13 Ta. In the current case of 1T on 1H, it leads to charge transfer from the T to the H and a depopulation of the flat band, since the H layer has a higher work function by 0.5 eV \cite{shimada_work_1994} (see also SI). For the monolayer 1T on graphene-based substrates, however, the charge transfer could be in the opposite sense as graphene has a much lower work function \cite{Mammadov_work_2017}, creating the expectation of substantial electron doping into the 1T layer. Such charge transfer effects are only now starting to be fully considered \cite{tilak_proximity_2024}, and we point out that spectroscopically there may be only a subtle distinction between a fully filled flat band due to complete charge transfer from the substrate, versus the lower of two Hubbard bands in a cluster Mott scenario \cite{gao_orbital_2025}. 

The metallic Fermi surface observed on the T terminations is relevant to the intriguing STM results on H-T bilayers. The metallic T states would contribute a narrow density of states close to $E_F$, which could provide an alternative understanding to the claim of heavy fermions \cite{vano_artificial_2021} or Kondo singlet formation \cite{wan_evidence_2023}. However, differing interlayer distances and charge transfer are important \cite{sanchez-ramirez_charge_2024}, therefore caution is required to translate our results and interpretation to the various different recent experimental configurations.

Our identification of complete charge transfer for the bulk of 4H\textsubscript{b} has clear implications for theoretical explanations of its intriguing superconductivity. With a charge transfer of 1 $e^-$ per 13 Ta from the T to the H layer, the T layers are thus effectively insulating, not because of the Mott-Hubbard correlations, but simply because charge transfer completely depopulates the flat band. As such, there are no itinerant T states that could couple to the superconductivity via interlayer hopping, nor local moments in 4H\textsubscript{b}-TaS\textsubscript{2} (consistent with the magnetic susceptibility \cite{di_salvo_preparation_1973}). This points against theoretical explanations that rely on magnetic moments or fluctuations in the T layers \cite{dentelski_robust_2021,luo_is_2024} and/or Kondo-like coupling \cite{persky_magnetic_2022,lin_kondo_2024}. Instead, our data instead supports the scenario of Josephson-like coupling between H layers through the insulating T layers \cite{fischer_mechanism_2023}. However, we point out that in the H layers there is also a competing low temperature \threebythree CDW; as such, there may be analogies to the case of NbSe\textsubscript{2} to further explore \cite{rahn_gaps_2012,flicker_charge_2016}. Our results on the anomalous termination also reinforce the fact that the crystals of 4H\textsubscript{b}-TaS\textsubscript{2} are imperfect, and stacking faults may be frequent, which should be taken into account when interpreting subtle experimental effects. 

Finally, our measurements underscore the robustness of the $\sqrt{13}$ CCDW order in the 1T layers to band filling and even local stacking orders. This is particularly interesting because, at cryogenic temperatures, the $\sqrt{13}$ CCDW, a \threebythree CDW (possibly incommensurate) and superconducting orders coexist \cite{friend_electrical_1977,ekvall_atomic_1997,nayak_evidence_2021}. A scenario of an underlying CDW with a superconducting one can make Higgs modes associated with the superconducting order accessible to spectroscopic probes \cite{shimano_higgs_2020}, as famously shown for 2H-NbSe\textsubscript{2} \cite{measson_amplitude_2014}. The CCDW in 4H\textsubscript{b}-TaS\textsubscript{2} also bears some analogies to the RTe\textsubscript{3} family, in the sense of the CDW breaking mirror symmetry, sometimes referred to as "ferroaxial" \cite{wang_axial_2022, alekseev_charge_2024}. In that context, an axial Higgs-like mode is proposed to arise purely from the CDW reconstruction that might couple with an external magnetic field \cite{wulferding_magnetic_2025}. This motivates the investigation of possible non-trivial orbital textures in 4Hb-TaS\textsubscript{2} \cite{ritschel_orbital_2015,alekseev_charge_2024}.  

In summary, our high-quality measurements of the electronic structure of 4H\textsubscript{b}-TaS\textsubscript{2} highlight the key role of charge transfer: the T termination has a metallic Fermi surface, but bulk T layers have complete charge transfer, emptying the flat band to leave effectively a band insulating T layers between the superconducting H layers. As well as clarifying that the interlayer coupling in the superconducting state must therefore be Josephson-like, our observation of the flat band at $E_F$ on the T termination may inspire reinterpretation of the zero-bias peaks observed in STM measurements of H-T bilayers. Our results highlight the benefit of spatial resolution in ARPES measurements, and will motivate other spatially resolved probes such as Raman spectroscopy and STM to provide further insights into the interplay of the CDW order and superconductivity, and potentially the signatures of Higgs modes, in the ever fascinating  4H\textsubscript{b}-TaS\textsubscript{2}. 

\section{Acknowledgments}
We thank T. K. Kim, B. Pal, R.~Valent\'{i}, S. Crampin and M. R\"{o}sner for stimulating discussions. We thank Diamond Light Source for access to beamline I05 under proposal SI36633. EDC acknowledges funding from the Royal Society through grant IES/R2/212016 and from UKRI grant UKRI597.

%\bibliography{ref_matt,ref_mihir}
%\bibliographystyle{naturemag}

%\bibliography{ref_matt}
%apsrev4-2.bst 2019-01-14 (MD) hand-edited version of apsrev4-1.bst
%Control: key (0)
%Control: author (8) initials jnrlst
%Control: editor formatted (1) identically to author
%Control: production of article title (0) allowed
%Control: page (0) single
%Control: year (1) truncated
%Control: production of eprint (0) enabled
%

%\clearpage

\section{Methods}
\subsection{Crystal growth}
4H\textsubscript{b}-TaS\textsubscript{2} crystals were grown by chemical vapour transport using iodine. Tantalum powder (99.9\%) and sulfur (99.99\%) were loaded in a quartz ampule together with anhydrous iodine. This procedure was performed in a nitrogen glovebox. The ampule was vacuum sealed under dynamic vacuum with a flame torch and placed in a two zone tube furnace with a temperature profile 800-850$^\circ$C. After 12 days it was cooled down with a rate of <12$^\circ$C/h.
\subsection{ARPES methods} 

ARPES measurements were performed on the nano-ARPES branch of beamline I05 \cite{hoesch_facility_2017} at Diamond Light Source. A capillary mirror optic was used to focus the light to a spot size of 4-5 $\mu{}$m (FWHM). Measurements were performed at a sample temperature of 25-30~K. The combined energy resolution was approximately 15 meV for the data measured with a photon energy of 50 eV (Fig 2 and Fig 3c,d,e) and 40 meV for the data at 100 eV (Fig 1 and Fig 3a,b). 

\subsection{Computational methods}

All first-principles calculations were carried out within density functional theory (DFT) using the VASP package and the projector-augmented wave formalism \cite{kresse_ultrasoft_1999}. Exchange-correlation energy functional parameterized by Perdew-Burke-Ernzerhof (PBE) \cite{perdew_generalized_1996} was employed, while van der Waals force was included through the DFT-D3 scheme with Becke-Johnson damping \cite{grimme_effect_2011}. Electron correlations on Ta-5d states belonging to the 1T sublayers were treated with Dudarev's DFT+$U$ approach \cite{dudarev_electron-energy-loss_1998} ($U_\textrm{Ta,5d}^\textrm{1T}=1.76$ eV \cite{ayani_unveiling_2024}).

To model the 4Hb stacking we used a $\sqrt{13}\times\sqrt{13}$ supercell for each 1T layer to capture the SoD CDW order, while keeping the adjacent 1H layers in their undistorted $1\times1$ form in the light of introducing a $3\times3$ CDW in 1H has only a marginal difference in interlayer charge transfer~\cite{crippa_heavy_2024}. Spin-orbit coupling was included in the Fermi surface calculation to describe the valley splittings at K and K' points originating from 1H layer~\cite{de_la_barrera_tuning_2018}, while omitted in band structure calculations that demonstrates negligible effects on the band structures along M-$\Gamma$ path, due to the weak interlayer coupling~\cite{almoalem_charge_2024} and its vertical mirror $\sigma_\textrm{v}$ inside. A plane wave cut-off energy of 400 eV and a $12\times12$ $\Gamma$-centered k-mesh were used together with a 30 meV Gaussian smearing. Electronic energy convergence criteria was set to $10^{-7}$ eV, and ions were relaxed until Hellmann-Feynman residual forces were below 1 meV/\AA{}. Band structure unfolding was carried out with VASPKIT code \cite{wang_vaspkit_2021}. The single-particle spectral function $A(k,\omega)$ shown in Fig.~\ref{Fig2}(d,g) was obtained by dressing the unfolded DFT eigenvalues with a Lorentzian broadening of $\Gamma$=0.05 eV:

\begin{equation*}
A(k,\omega)=\sum_{n} w_n(k)\frac{1}{\pi}\frac{\Gamma}{[\omega-\varepsilon_n(k)]^2+\Gamma^2}
\end{equation*}

where $\varepsilon_n(k)$ denotes an eigenvalue of $n$-th band at a given momentum $k$ in the unfolded 1st Brillouin zone, and $w_n(k)$ is the unfolding weight. Using a momentum-independent $\Gamma$ amounts to assuming a uniform quasiparticle lifetime and neglecting matrix-element effects, which is sufficient for the qualitative comparison presented here.

% Author contributions - tbc later

\end{document}